\documentclass[twocolumn,prb]{revtex4}%
\usepackage{epsfig}
\usepackage{amsmath}
\usepackage{amsfonts}
\usepackage{amssymb}
\usepackage{graphicx}%
\begin{document}
\title{Strain selection of charge and orbital ordering patterns in half-doped
manganites}
\author{M.J. Calder\'on}
\affiliation{Cavendish Laboratory, Cambridge University, Madingley Road,
Cambridge CB3 0HE, UK}
\author{A.J. Millis}
\affiliation{Department of Physics, Columbia University, 538 W. 120th St, New York, NY 10027}
\author{K.H. Ahn}
\affiliation{Theoretical Division, Los Alamos National Laboratory, Los Alamos,
New Mexico 87545}
\date{\today}

\begin{abstract}
Theoretical and computational results are presented clarifying the role of
long-ranged strain interactions in determining the charge and orbital ordering
in colossal magnetoresistance manganites. The strain energy contribution is
found to be of order $20-30$ $meV/Mn$ and in particular stabilizes the
anomalous 'zig-zag chain' order observed in many half-doped manganites.

\end{abstract}
\maketitle

\begin{figure}[ptb]
\epsfxsize=2.in
\epsffile{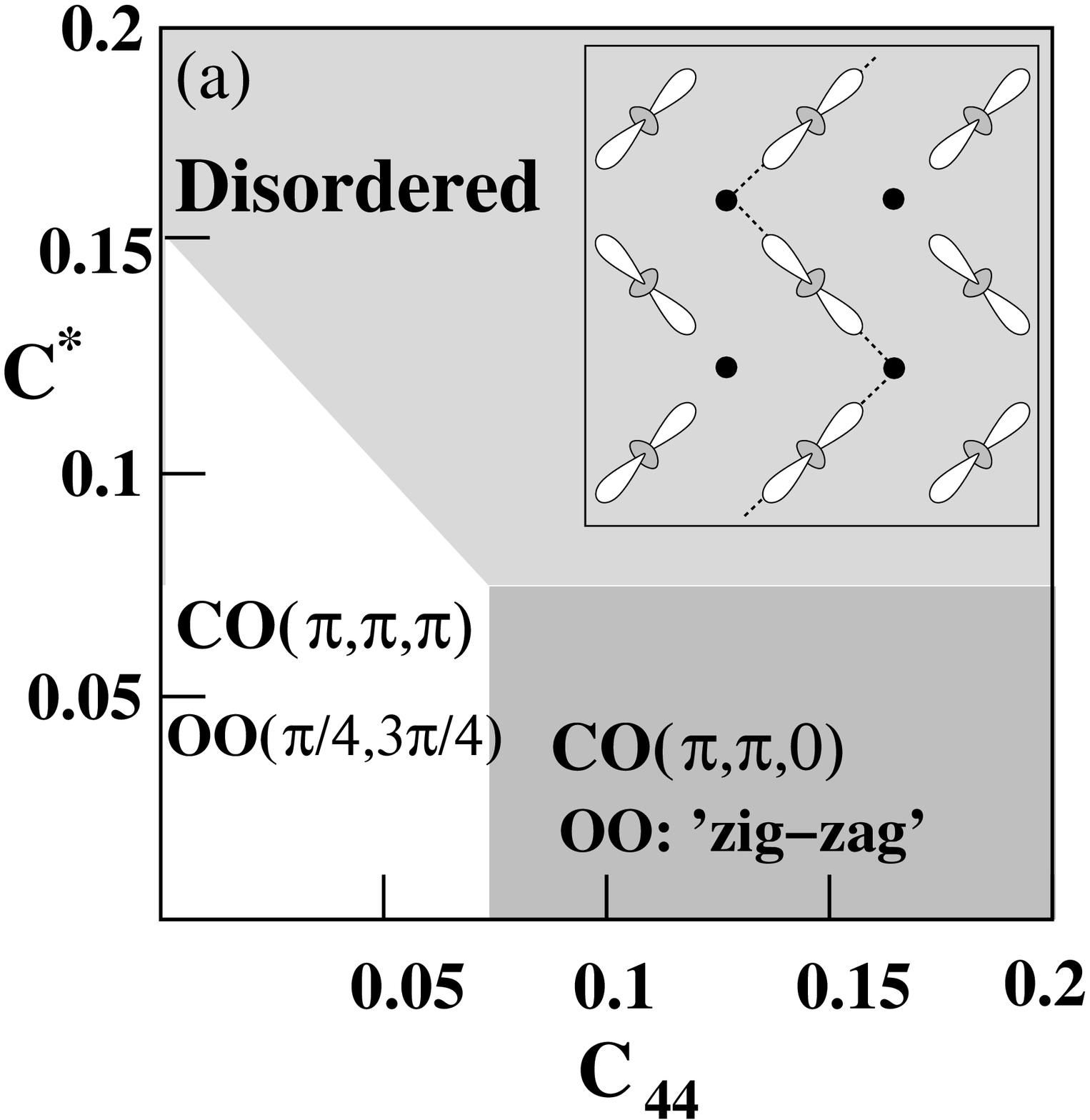}
\epsfxsize=2.in
\epsffile{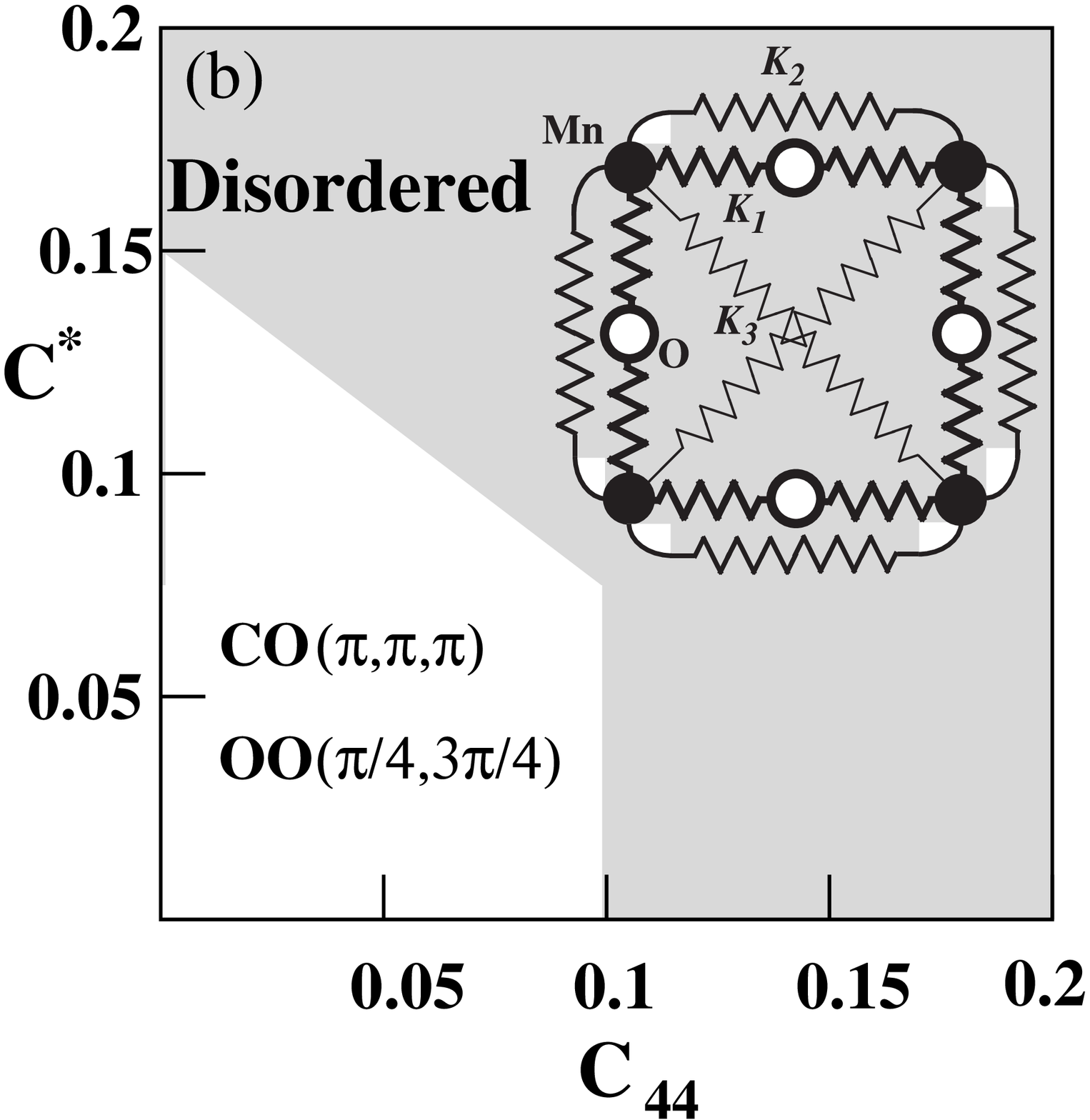}
\caption{ Main panels: ground-state phase diagrams calculated by Monte-Carlo
annealing on $8\times8\times8$ lattices, with (a) and without (b) the inclusion
of the
long range energy. The wave vector of the charge order (CO) and the pattern
$\theta_{i}$ of the orbital ordering (OO) are indicated. 'Disordered'
corresponds to phases which show no obvious periodicity or symmetry on the
lattices studied. Calculations are performed for $\beta=0.8$ for various force
constants consistent with bulk modulus $C_{B}=0.35K_{1}$ and are presented as
functions of the shear $C_{44}$ and Jahn-Teller $C^{*}$ elastic moduli implied
by the force constants and expressed in units of $K_{1}$. The inset to Fig.
\ref{fig:pd}a
shows the 'zig-zag' pattern. Black dots represent the sites with low charge
occupancy. (b)The 'zig-zag' pattern is absent when uniform strain energy is
neglected. The inset to Fig. \ref{fig:pd}b shows three
of the force constant model couplings. Additional non-central force coupling to Mn-Mn-Mn bond angles not shown. }%
\label{fig:pd}%
\end{figure}

Perovskite manganites of chemical formula Ln$_{1-x}$Ak$_{x}$MnO$_{3}$ (Ln is a
lanthanide rare earth such as La or Pr; Ak is a divalent alkali such as Sr or
Ca) present a complex phase diagram associated with the interplay of charge,
spin, orbital and lattice degrees of freedom. \cite{reviews} Charge and
orbitally ordered phases appear in wide parameter ranges, for example in
La$_{1-x}$Ca$_{x}$MnO$_{3}$ for $x\geq0.5$, and in Pr$_{1-x}$Ca$_{x}$MnO$_{3}$
for $0.3\leq x\leq0.7$. The physical mechanism underlying the ordering remains
the subject of debate. Particular attention has focussed on the half-doped
($x=0.5$) materials, including La$_{0.5}$Ca$_{0.5}$MnO$_{3}$, Pr$_{0.5}%
$Ca$_{0.5}$MnO$_{3}$ and Nd$_{0.5}$Sr$_{0.5}$MnO$_{3}$, which possess a
strongly insulating ground state with a particularly interesting ('zig-zag
chain') spin, charge and orbital ordering pattern shown in the inset to
Fig. \ref{fig:pd}a. Stabilizing this state requires \ 'second neighbor'
interactions, which as noted by Khomskii and Kugel \cite{khomskii01} and by
Ahn and Millis \cite{Ahn01} may reasonably be expected to arise from
electron-lattice interactions. However, experiments have also indicated that
\textit{long ranged strain} plays an important role. Insulating ordered phases
appearing in bulk La$_{0.5}$Ca$_{0.5}$MnO$_{3}$ are suppressed in thin films
\cite{nyeanchi99}; it is argued that the difference arises from the lattice
mismatch with the substrate, which prevents the occurrence of lattice
distortions necessary to accommodate orbital ordering. Similar conclusions have
been drawn from recent work on Pr$_{0.6}$Ca$_{0.4}$MnO$_{3}$ films subjected
both to compressive and tensile strain. \cite{nelson03} Other sets of
experiments in polycrystalline La$_{0.5}$Ca$_{0.5}$MnO$_{3}$ show that the
charge/orbital ordering transition temperature is progressively suppressed as
grain size is decreased; the interpretation again is that boundary effects
inhibit the formation, in small grains, of the strains imposed by the orbital
ordering. \cite{levy00} Perhaps most significantly, polarized optical
microscopy studies on Bi$_{0.2}$Ca$_{0.8}$MnO$_{3}$ single crystals
\cite{podzorov01} reveal that as temperature is decreased to just below the
charge/orbital ordering transition, lenticular shaped domains appear and grow
slowly with time. This behavior is well known in martensitic systems
\cite{martensite}, where it is attributed to the interplay between a tendency
of the system to favor a state with large strain, and a boundary condition
which prevents the existence of a truly uniform strain.

Despite the considerable experimental evidence that long ranged strain is
important, strain physics has not received much theoretical attention. The
original analyses \cite{goodenough,kanamori,kugel} and many subsequent works
\cite{theory,millis96,yunoki00} focused on local interactions, including
magnetic, Coulomb and Jahn-Teller electron-phonon coupling. In this paper
we study long ranged strain effects theoretically, and in particular show that
they play a crucial role in stabilizing the particular ordered state
characteristic of many of the half doped materials.

The Ln$_{1-x}$Ak$_{x}$MnO$_{3}$ materials have approximately the cubic
perovskite crystal structure, in which the Mn ions sit on the vertices of a
simple cubic lattice. The important low-energy electronic degrees of freedom
may be thought of as Mn $e_{g}$ electrons. There are two $e_{g}$ orbitals per
Mn. These are degenerate in cubic symmetry. The ($1-x$) $e_{g}$ electrons are
relatively mobile at room temperature but in many manganite materials,
including the half-doped compounds of present interest, they become strongly
localized at low temperatures. The $x\neq0$ insulating states have a charge
density which varies from Mn to Mn; the pattern of charge density is referred
to as 'charge order'. In many of the half-doped materials, periodic charge
order is observed, with wave vector $(\pi,\pi,0)$: in other words, in a given
plane (which we take to be the $xy$ plane) of the ideal cubic perovskite
structure the charge alternates from site to site in a simple two sublattice
fashion, but the charge ordering simply stacks uniformly in the perpendicular
direction. The observation of $(\pi,\pi,0)$ is surprising: charge ordering
typically occurs because electrons repel each other, and a fully alternating
$(\pi,\pi,\pi)$ ordering is typically preferred because it allows the greatest
distance between charges.

If charges are localized, then one expects that on a given site one linear
combination of the two $e_{g}$ orbitals will have a higher occupancy than the
orthogonal linear combination. The pattern of this preferential occupancy is
referred to as 'orbital ordering'. It was first inferred from the magnetic
ordering pattern (discussed below) by Wollan and Koehler in the 1950s
[\onlinecite{wollan}] using rules determined by Goodenough \cite{goodenough}
and Kanamori \cite{kanamori}, and has now been studied in numerous samples by
resonant x-ray scattering \cite{ishihara02}, which is sensitive both to the
orbital occupancy and to the changes it induces in oxygen positions. The
orbital ordering occurring in the strongly insulating half doped materials may
be visualized as 'zig-zag chains' on the $xy$-plane stacked uniformly in the
$z$-direction. The 'zig-zag' is produced by the alternation of preferred
occupancy of $3x^{2}-r^{2}$ and $3y^{2}-r^{2}$ orbitals on the sublattice of
sites with high charge occupancy (see inset to Fig. \ref{fig:pd}a). The
magnetic ordering pattern corresponds to ferromagnetism along the 'zig-zag
chains' but with spin direction alternating from chain to chain, and follows
from the application of the Goodenough/Kanamori rules to the zig-zag pattern
of states \cite{kugel}. We note, however, that strain physics should also be
important because the $e_{g}$ orbitals have a non-cubic shape. Thus, in the
case of 'zig-zag' ordering, the occupied orbitals have charge density lying
mainly in the 'xy' plane, leading to a compression in the perpendicular
direction.

We now turn to the theory. A general model would include electronic dynamics
(hopping and electron-electron interaction), electron-lattice coupling, and
the energetics associated with the lattice degrees of freedom. Because we wish
to focus on the new physics associated with strain effects, and because the
main application we have in mind is to strongly insulating half-doped
manganites, for simplicity we adopt a classical, insulating model in which the
electrons are taken to be localized on Mn sites, but our approach may easily
be extended to an itinerant electron model. In the insulating model the
e$_{g}$ orbitals of a Mn site can be either empty, or occupied by one electron
which resides in a linear combination of the two $e_{g}$ orbitals. We
parameterize the charge state of site $i$ by a variable $h_{i}$ with $h_{i}=0$
for an occupied site and $h_{i}=1$ for an empty site, so doping $x=\left(
\sum_{i}h_{i}\right)  /N$ and we focus here on $x=0.5$. The orbital state of
an occupied site is described by an angle $0\leq\theta_{i}\leq\pi$ so that
$|\psi_{i}(\theta_{i})\rangle=\cos{\theta_{i}}|d_{3z^{2}-r^{2}}\rangle
+\sin{\theta_{i}}|d_{x^{2}-y^{2}}\rangle.$ (Note that $|d_{3z^{2}-r^{2}
}\rangle$, $|d_{3x^{2}-r^{2}}\rangle$, $|d_{3y^{2}-r^{2}}\rangle$ correspond to
$\theta_{i}
=0,\pi/3,2\pi/3$ respectively while $|d_{z^{2}-x^{2}}\rangle$,
$|d_{x^{2}-y^{2}}\rangle$,
$|d_{z^{2}-y^{2}}\rangle$, correspond to $\theta_{i}=\pi/6,\pi/2,5\pi/6$.)

The electron-lattice coupling is taken to have the form given in Ref.
[\onlinecite{millis96}]: a site containing an electron is Jahn-Teller coupled
to the surrounding oxygen octahedron and an empty site induces a breathing
distortion of the octahedron. In an isolated MnO$_{6}$ octahedron the
Jahn-Teller effect would lead to an energy gain $E_{JT}=\lambda^{2}/2K_{1}$
with $\lambda$ the Jahn-Teller coupling and $K_{1}$ the Mn-O bond-stretching
force constant (see the inset to Fig. \ref{fig:pd}b). Optical measurements on
LaMnO$_{3}$
suggest $K_{1}\approx7.36eV/\mathring{A}^{2}$ and both band structure
calculations and fitting to the observed structure suggest $\lambda
\approx1.38eV/\mathring{A}$ so that $E_{JT}\approx0.26eV$. We adopt this as
the basic energy scale of the theory. We write the energy gain due to a
breathing distortion as $E_{B}=\beta E_{JT}$. We expect that the energy costs
of the two distortions are similar, so $\beta \sim 1$.

In the actual perovskite lattice, each O ion is bonded strongly to two Mn, so
a breathing or Jahn-Teller distortion about one site is coupled to distortions
at neighboring sites. To incorporate this physics we follow Refs.
[\onlinecite{millis96,ahn98}] and write a model whose variables are the
deviations $\vec{\delta}_{i}$ of the Mn ions from their ideal cubic
perovskite positions and the displacements $\vec u_i$ of the O-ion
at $\vec{i}+\hat{a}/2$ along $\hat{a}$ direction
($a=x,y,z$).  These variables are coupled harmonically
by near neighbor force constants. The
inset to  Fig. \ref{fig:pd}b shows the planar force constants. We parameterize
these force
constants by the bulk ($C_{B})$, cubic-tetragonal ($C^{\ast}$) and shear
$(C_{44}$) moduli which they imply. We then integrate out the variables
describing the O positions and obtain an energy of the form
\begin{equation}
E_{\mathrm{tot}}(\{h_{i},\theta_{i},\delta_{i}\})=E_{\text{lattice}}
(\{\delta_{i}\})+E_{\mathrm{JT,breath}}(\{h_{i},\theta_{i},\delta_{i}\})
\label{eq:TOT}%
\end{equation}
(A detailed derivation will be given in a separate communication \cite{long}).
\

For each electronic configuration $(\{h_{i},\theta_{i}\})$ we may now obtain
the total energy by minimizing Eq. \ref{eq:TOT} with respect to the lattice
variables $\vec{\delta}_{i}$. We have used a classical Monte Carlo
simulation process on $4\times4\times4$ and $8\times8\times8$ Mn lattices to
find
the ground state by annealing the system starting from a random high
temperature configuration. However, an issue arises in the evaluation of the
energy. To describe this, note that the term $E_{\mathrm{JT,breath}}$ may be
conveniently written in momentum space as a linear coupling between the
Fourier transform of the displacements, $\vec{\delta}_{k}$, and a
vector $\vec{e}_{k}$ constructed from the $h_{i},\theta_{i}$ and
the parameter $\beta$: $E_{\mathrm{JT,breath}}=\sum_{k}\vec{e}%
_{-k}\mathbf{\Lambda}(k)\vec{\delta}_{k}$. The purely lattice term
may similarly be written as $E_{\text{lattice}}=\frac{1}{2}\sum_{k}%
\vec{\delta}_{-k}\mathbf{K}(k)\vec{\delta}_{k}$ with $\mathbf{K}$
a force-constant matrix, so that after minimization over the $\delta$ we have%
\begin{equation}
E_{\mathrm{tot}}=-\frac{1}{2}\sum_{k}\vec{e}_{-k}\left[
\mathbf{\Lambda}(k)\mathbf{K}^{-1}(k)\mathbf{\Lambda}(k)\right]
\vec{e}_{k} \label{Ek}%
\end{equation}

On a finite lattice the sum is over a discrete set of $k$ points and the terms
with $k\neq0$ may straightforwardly be evaluated. However, as $k\rightarrow0$
we find $\mathbf{\Lambda}(k)\sim k$ and $\mathbf{K}(k)\sim k^{2}$ so that the
$\lim_{k\rightarrow0}\left[  \mathbf{\Lambda}(k)\mathbf{K}^{-1}%
(k)\mathbf{\Lambda}(k)\right]  \sim\frac{0}{0}$. The limit, and therefore the
value of the $k=0$ term in Eq. \ref{Ek} depends on the \textit{direction} from
which $k$ approaches $0.$ These different limits correspond to different
uniform strain states of the material (and therefore to different changes to
shape of the whole lattice), and require special attention. We find that for
almost every configuration of $h_{i},\theta_{i}$ there is one particular
strain state (direction along which the $k\rightarrow0$ limit is taken) which
minimizes $E_{tot}$. We use two annealing procedures. In one, we use the
$E_{tot}$ corresponding to this 'optimum' strain state, thereby updating the
strain every time the electronic configuration is changed and allowing the
system to minimize its energy with respect to strain. \cite{note} In the
other, we neglect the $k=0$ term in Eq. \ref{Ek} entirely. One could also
constrain the $k=0$ Fourier components of $\delta_{i}$ to take particular
values, thereby forcing the system to accommodate to a definite strain (to
simulate, e.g. a film grown on a lattice-mismatched substrate), but we do not
explore this physics here.

We have conducted extensive simulations of Eq. \ref{eq:TOT} for different
parameters. The bulk modulus has been estimated to be $135$ GPa ($0.35$ in
units of K$_{1}$) \cite{bulkmodulus} and we consider only choices of $K_{i}$
consistent with this value. We present our results (obtained by varying the
$K_{i}$ in a manner consistent with the experimental bulk modulus) in terms of
the value of the other elastic moduli. Different values for $\beta$ have been
explored, but here we only show results for $\beta=0.8$.

\begin{table*}[ptb]
\caption{Energies for different orbital (OO) and charge ordering (CO) patterns
as indicated, calculated for parameters $\beta=0.8$, $C_{B}=135$ GPa ($0.35
K_{1}$), $C^{*}=17.35$ GPa ($0.045 K_{1}$), and $C_{44}=30.85$ GPa ($0.08
K_{1}$) chosen within the range in which the 'zig-zag' ordering is the ground
state (see Fig.\ref{fig:pd}a).}%
\label{table:energies}
%
\begin{tabular}
[c]{c|c|c|c}
\hline
& Uniform Strain
Energy (meV/Mn)
& Short Range
Energy (meV/Mn) &
Total
Energy (meV/Mn)\\\hline
CO($\pi,\pi,0$), OO 'zig-zag' & -33.2 & -278.4 & -311.6\\
CO($\pi,\pi,\pi$), OO 'zig-zag' & -33.2 & -270.9 & -304.2\\
CO($\pi,\pi,0$), OO($\pi/2,\pi/2$) & -65.0 & -218.0 & -283.0\\
CO($\pi,\pi,\pi$), OO($\pi/4,3\pi/4$) & -15.9 & -290.8 & -306.7\\\hline
\end{tabular}
\end{table*}

Fig. \ref{fig:pd} shows typical ground-state phase diagrams arising from our
simulations. The main panel in Fig. \ref{fig:pd}a shows the phase diagram following from
optimizing the full model over strain. Three states are observed: one
($CO=(\pi,\pi,\pi)$; $OO=(\pi/4,3\pi/4)$) corresponds to the arrangement of
charges and orbitals which implies the lowest uniform strain; another,
labeled 'Disordered' in the figure, corresponds to states which have no
obvious ordering pattern within the $8\times8\times8$ unit cells we can access
numerically, and which presumably either have some extremely long-ranged
periodicity or are glassy. We note that in this parameter regime these
'disordered' states are found by the annealing procedure to have energies
lower than those of any simple periodic structure we have devised.

In the lower right region of the phase diagram in Fig. \ref{fig:pd}a the
preferred state corresponds to the one found in experiments. Because it
involves quite substantial anisotropic lattice distortions it is stabilized
only in particular regions of the phase diagram. The main panel of Fig. \ref{fig:pd}b  shows
the results of computations for the same parameters, but with the uniform
strain ($k=0$) contribution to the energy neglected. We see that the
experimentally observed phase is entirely absent. This conclusion
remains true in the range $0.6\leq\beta\leq1.2$. Details on the role of $\beta$
will be given in a
separate communication.\cite{long}

The difference in phase diagrams computed with and without the strain
contribution to the energy suggests the importance of long ranged strain
effects. To understand this result in more detail we have evaluated both the
long range ($k=0$) and short range ($k\neq0$) contributions to $E_{tot}$ for
many different orbital orders, and $(\pi,\pi,0)$ and $(\pi,\pi,\pi)$ charge
order. Typical results are presented in Table I. One sees that the differences
in strain energy between different orbitally ordered phases ($\sim20meV/Mn)$
are of the same order as magnetic energy differences computed by other workers
\cite{yunoki00}. Observe that the uniform strain energy only depends on the
orbital configuration, and is optimized for orbitals which correspond to
mainly planar charge densities. The 'zig-zag' pattern of orbitals is
consistent with this result. The 'uniform clover' pattern (all orbitals
$\theta=\pi/2$ i.e. $d_{x^{2}-y^{2}}$) has an even lower strain energy, but in
the insulating model considered here its short range energy is never large
enough to stabilize this phase. We note, however, that electron banding
effects are found theoretically to favor this phase \cite{maezono98} and that
it is observed in metallic La$_{0.5}$Sr$_{0.5}$MnO$_{3}$. On the contrary, the
$(\pi/4,3\pi/4)$ orbital state is the least strained configuration of all the
ordered ones, the reason for this being its isotropic charge distribution.
Note also that the strain term does not distinguish
between different z-direction orderings: it is the short range part of the
energy which chooses the stacking of charge in the z-direction ($(\pi,\pi,0)$
vs $(\pi,\pi,\pi)$).

We now comment on possible extensions of our model. Magnetic interactions have
been neglected, but could be easily included. The $t_{2g}$ localized spins are
antiferromagnetically coupled via superexchange. There is also a ferromagnetic
superexchange interaction between the $e_{g}$ spins which depends on the
occupied orbitals in two neighboring sites. \cite{millis97} If applied to our
'zig-zag' orbital configuration, CE-type antiferromagnetism is found. Magnetic
energy differences per site between CE-type antiferromagnetism and
ferromagnetic ordering are $\sim20meV$, therefore of the same order than the
uniform strain energies. Yunoki and co-workers \cite{yunoki00} have shown how
the magnetic interaction may help to stabilize $(\pi,\pi,0)$ ordering relative
to $(\pi,\pi,\pi)$ ordering; the energetics found in their work are
sufficiently similar to those found by us that it is not clear which mechanism
is dominant. Similarly, we have adopted a strictly classical, insulating model
for the electronic degrees of freedom, but our electron-lattice coupling and
lattice interaction could be easily carried over to an itinerant electron
model (at least within the conventional adiabatic approximation for
electron-lattice coupling). Use of such a model to explore strain effects on a
second order weak amplitude ('Charge Density Wave' instability) electronic
charge and orbital ordering transition would be very interesting. Finally, our
calculation involves only relatively small length scales, and therefore has
not included elastic compatibility and other long range effects. We hope our
results may be useful as input parameters for calculations of longer length
scale phenomena.

In summary, a classical model for charge and orbital ordering has been
formulated and analyzed. The important new feature is that elastic energies as
well as the cooperative Jahn-Teller coupling are included. Our main finding is
that energetics associated with long ranged strain are at least as important
as those from other (e.g. magnetic) sources in determining the ground state of
the system. We find in particular that uniform strain is of vital importance
in stabilizing the 'zig-zag chain' orbital ordering pattern observed in many
manganite systems.

\emph{Acknowledgments.} MJC acknowledges financial support from Churchill
College (Cambridge, UK), AJM from the University of Maryland-Rutgers MRSEC
(NSF-DMR-00080008), and KHA from US DOE.


\begin{thebibliography}{99}
\bibitem {note}Note that, in reality, a whole single crystal cannot
dramatically change its dimensions due to the internal strain but rather
different domains where strain sets in different directions would be expected.
Such martensitic domains are typically much larger than the sample-sizes
numerically accessible to us, so we consider only single-domain behavior.
\end{thebibliography}
\end{document}